\documentclass[aps,prd,preprintnumbers,showpacs]{revtex4}
\usepackage[dvips]{graphicx}

\begin{document}

%
%

\eprint{Nisho-1-2020}
\title{A New Method for Detecting Axion With Cylindrical Superconductor }
\author{Aiichi Iwazaki}
\affiliation{International Economics and Politics, Nishogakusha University,\\ 
6-16 3-bantyo Chiyoda Tokyo 102-8336, Japan.}   
\date{July. 20, 2020}
\begin{abstract}
We propose a method for searching dark matter axion in axion-photon conversion.
We consider a superconductor of cylindrical shape under strong magnetic field.
The dark matter axion generates oscillating electric field which induces
oscillating superconducting current in the surface of the superconductor.
The current gives rise to dipole radiation with the frequency $m_a/2\pi$ given by axion mass $m_a$.
We show that the radiation flux generated by the
current is of the order of $10^{-18}$W under magnetic field $\sim 3$T in the case of  
radius $1$cm and length $10$cm of the cylindrical superconductor. 
The large amount of the radiation flux arises because Cooper pairs with large 
number density ( $\sim 10^{22}/\rm cm^{3}$ ) is present in the superconductor.
With the high detection sensitivity,
we can simultaneously search wide bandwidth of the radio frequency with existing radio telescope.
\end{abstract}
\hspace*{0.3cm}

\hspace*{1cm}

\maketitle

Axion has been introduced \cite{axion} for solving strong CP problem in QCD.
The axion is the Nambu-Goldstone boson\cite{axion} associated with U(1) Peccei-Quinn symmetry. 
The symmetry is chiral and naturally solves the strong CP problem: The problem is why the CP violating term $G_{\mu,\nu}\tilde{G}^{\mu,\nu}$ is
absent in QCD Lagrangian where $G_{\mu,\nu}$ ( $\tilde{G}^{\mu,\nu}$ ) denotes ( dual ) fields strength of gluons.
The axion is the phase of a complex scalar field carrying the U(1) charge.
Although the axion is the Nambu-Goldstone boson, it acquires 
its mass $m_a$ through chiral anomaly because the Peccei-Quinn symmetry is chiral. 
Thus, instantons in QCD give rise to the mass of the axion. The axion is called as QCD axion.
In the paper we mainly consider the QCD axion. The axion is generated in early universe 
and could become one of dominant components of dark matter in the present universe when the Peccei-Quinn symmetry
is broken after inflation\cite{Wil}.

\vspace{0.1cm}
In the present day, the axion is one of the most promising candidates for the dark matter.
There proceed several projects for the detection of the axion. There are three types of the projects; Haloscope, Helioscope and others.
The dark matter axion in our galaxy is searched with Haloscope, while axion produced in the Sun is searched with Helioscope.
Haloscope projects are ADMX\cite{admx}, CARRACK\cite{carrack}, HAYSTAC\cite{haystac}, 
ABRACADABRA\cite{abracadabra}, ORGAN\cite{organ}, etc. Helioscope projects are CAST\cite{cast}, SUMICO\cite{sumico},  etc..
The others are LSW\cite{lsw} ( Light Shining through a Wall ), VMB\cite{vmb} ( Vacuum Magnetic Birefringence ) etc.
There are many axion-like particles proposed other than the QCD axion. Thus, 
the mass range in the search is very wide, $10^{-20}\rm eV$$\sim 1\rm eV$ or more \cite{obs}. 
But we assume the QCD axion as the dominant component of the dark matter
and expect that the appropriate mass range is of the order of $10^{-6}\rm eV$$\sim 10^{-4}\rm eV$,
as suggested by cosmological models and lattice gauge theories\cite{Wil, lattice}.

\vspace{0.1cm}
In previous papers\cite{iwazakifrb} we have explained origin of fast radio bursts ( FRB) assuming the QCD axion. The characteristic features of the FRBs\cite{frb} 
are their frequency $\sim 1$GHz and very short duration $\sim 10^{-3}$s. In our model 
the bursts arise in the collision of axion star\cite{axionstar} and neutron star, or magnetized 
accretion disk around black hole. ( The axion star causing the FRBs is gravitationally loosely bound state of the axions. )
The radiations are photons converted from axions under strong magnetic fields of these astrophysical objects.
We have obtained\cite{iwazakifrb2} the axion mass $\simeq 7\times 10^{-6}\rm eV$ by examining spectral feature of the repeating FRB121102\cite{frb121102}. 
Our proposal for the detection of the dark matter axion is mainly focused to such a mass range, 
although it can be applied to wider mass range than that we focus.

\vspace{0.1cm}
Our proposal is a type of Haloscope.
The point in our proposal is that electric field induced by dark matter axion under magnetic field generates oscillating electric current
in the surface of superconductor. The current is superconducting and 
produces the large amount of dipole radiations. This is because Cooper pairs with the large number density are present as carriers of the superconducting current.
Our system is composed of a superconductor of cylindrical shape and a receiver for the detection of the radiations.
We do not need to use resonant cavity. We need only sensitive radio receiver like radio telescope. 
We shall show in the paper that the larger amount of the radiation flux is generated by the cylindrical 
superconductor than those generated in resonant cavities
of Haloscope experiments.  We calculate the ratio of signal to noise and find that the detection sensitivity is very high.
We explain why such large amount of radiation flux is obtained in our 
apparatus, compared with resonant cavity experiments.

\vspace{0.1cm}

First we show that the coherent axion induces an electric field under a magnetic field.
It is well known that
the axion $a(\vec{x},t)$ couples with both electric $\vec{E}$ and magnetic fields $\vec{B}$ in the following,

\begin{equation}
\label{L}
L_{aEB}=k_a\alpha \frac{a(\vec{x},t)\vec{E}\cdot\vec{B}}{f_a\pi}
\end{equation}
with the decay constant $f_a$ of the axion
and the fine structure constant $\alpha\simeq 1/137$,   
where the numerical constant $k_a$ depends on axion models; typically it is of the order of one.
The standard notation $g_{a\gamma\gamma}$ is such that $g_{a\gamma\gamma}=k_a\alpha/f_a\pi\simeq 0.14(m_a/\rm GeV^2)$
for DFSZ model\cite{dfsz} and $g_{a\gamma\gamma}\simeq -0.39(m_a/\rm GeV^2)$ for KSVZ model\cite{ksvz}.
In other words, $k_a\simeq 0.37$ for DFSZ and $k_a\simeq -0.96$ for KSVZ.
The axion decay constant $f_a$ is related with the axion mass $m_a$ in the QCD axion; $m_af_a\simeq 6\times 10^{-6}\rm eV\times 10^{12}$GeV.
It should be mentioned that the discussions below can be also applied even for axion-like particles, which couple with the
electromagnetic fields in the way just as in eq(\ref{L}). 

\vspace{0.1cm}
The interaction term in eq(\ref{L}) slightly modifies Maxwell equations,

\begin{eqnarray}
\vec{\partial}\cdot\vec{E}+\frac{k_a\alpha\vec{\partial}\cdot(a(\vec{x},t)\vec{B})}{f_a\pi}&=0&, \quad 
\vec{\partial}\times \Big(\vec{B}-\frac{k_a\alpha a(\vec{x},t)\vec{E}}{f_a\pi}\Big)-
\partial_t\Big(\vec{E}+\frac{k_a\alpha a(\vec{x},t)\vec{B}}{f_a\pi}\Big)=0,  \nonumber  \\
\vec{\partial}\cdot\vec{B}&=0&, \quad \vec{\partial}\times \vec{E}+\partial_t \vec{B}=0.
\end{eqnarray}
From the equations, we approximately obtain the electric field $\vec{E}$
generated by the axion $a$ under the background magnetic field $\vec{B}$,

\begin{equation}
\label{E}
\vec{E}_a(r,t)=-k_a\alpha \frac{a(\vec{x},t)\vec{B}(\vec{x})}{f_a\pi}
\end{equation}
by assuming the parameter $k_a\alpha a/f_a$ extremely small and small momenta $p_a$ of the dark matter axion; $p_a\ll m_a$.
Typically, it is supposed to be given by $p_a\sim 10^{-3}m_a$ because of the nonzero velocity $v_a\sim 10^{-3}$ of the dark matter axion in our galaxy. 

\vspace{0.1cm}
The electric field $\vec{E}_a$ is the one obtained in the vacuum. 
In general the electric field induced by the axion is suppressed\cite{supp} depending on the medium or
the configuration of the magnetic field $B(\vec{x})$. Here we use superconductor.
The magnetic field penetrates the superconductor to the London penetration depth $\lambda \sim 10^{-5}$cm.
It strongly depends on the spatial distance $x$ from the surface; $B(x)\propto \exp(-x/\lambda)$.
It apparently seems that
there presents a suppression factor such as $(m_a\lambda)^2\ll 1$, as has been shown\cite{supp}.
But we can show that the suppression factor is absent in the superconductor described by Ginzburg-Landau model.
( We have solved the equations of electromagnetic fields coupled with axion and Cooper pair described by Ginzburg-Landau model. We have found that
the electric field in the superconductor is just given by $E_a$ in eq(\ref{E}). The paper about the discussion
is now in preparation. )

\vspace{0.1cm}
The energy density of the dark matter axion is given by

\begin{equation}
\rho_a= \frac{1}{2}(\dot{a}^2+(\vec{\partial} a)^2+m_a^2a^2)\simeq m_a^2a^2
\end{equation}
where $a(t)=a_0\cos(t\sqrt{m_a^2+p_a^2} )\simeq a_0\cos(m_a t)$.
The local energy density in our galaxy is supposed such as $\rho_a\simeq 0.3\rm GeV\,\rm cm^{-3}\simeq 2.4\times 10^{-42}\rm GeV^4$.
Assuming that the whole of the dark matter energy density comes from the dark matter axion,  
we obtain extremely small parameter $a/f_a\simeq \sqrt{\rho_a}/(m_af_a)\sim 10^{-19}$.
The energy density also gives the large number density of
the axions $\rho_a/m_a\sim 10^{15}\mbox{cm}^{-3}(10^{-6}\mbox{eV}/m_a)$, which causes their coherence. Thus we may treat the axions as the classical axion field $a$.

\vspace{0.1cm}
The Cooper pair in a superconductor under the electric field $E_a(t)$ oscillates with the frequency $m_a/2\pi$
according to the equation of motion, $m_c\dot{v}=qE_a$, where $m_c=2m_e$ ( $q=2e$ ) denotes the mass ( electric charge ) of the Cooper pair 
with electron mass $m_e$ and charge $e$.
We note that the motion of the Cooper pair is not disturbed by impurities in the superconductor,
so there is no dissipative term in the equation of motion.
Thus, the superconducting current density is given by $J=qnv=\frac{q^2E_a n}{m_am_c}$ 
with the number density $n$ of the Cooper pair; $n\sim 10^{22}\rm cm^{-3}$. 
The current oscillates with the frequency $m_a/2\pi$.
The frequency spectrum of the axions has the peak frequency $m_a/2\pi$
with small bandwidth $\Delta\omega\simeq 10^{-6}\times m_a$, because of nonzero velocity $v_a\sim 10^{-3}$ in our galaxy. 
Thus, the oscillating current has the same spectrum as that of the axion.

We impose the magnetic field $\vec{B}$ on a cylindrical superconductor, which is parallel to the direction along the length of the superconductor. 
We take the direction as $z$ direction. The magnetic field is expelled from the superconductor. But
the field penetrates into the superconductor to the depth $\lambda=\sqrt{m_c/q^2n}$ ( London penetration depth ).
The penetration depth is assumed such that $\lambda\simeq 5\times 10^{-6}\rm cm$ corresponding to the number density $n \simeq 0.5\times 10^{22}\rm cm^{-3}$.
Thus, the superconducting current flows in the surface to the depth $\lambda$. 
Since it oscillates with the frequency $\simeq m_a/2\pi$, the dipole radiation is emitted from the cylindrical superconductor.

\vspace{0.1cm}
Now, we estimate the radiation flux emitted by the cylindrical superconductor under the magnetic field $B$.
We suppose that the superconductor has radius $R=1$cm and length $l=10$cm.
Then,  the flux of the dipole radiation is given by 

\begin{equation}
\label{S}
S=\frac{m_a^2 (2\pi Rl\lambda J)^2}{3}=\frac{m_a^2(q^2E_0 n)^2(2\pi Rl\lambda)^2}{3m_a^2m_c^2}
=\frac{(q^2 k_a\alpha n B)^2(2\pi Rl\lambda)^2\rho_a}{3\pi^2m_a^2f_a^2m_c^2}=\frac{(k_a\alpha B)^2(2\pi Rl)^2\rho_a}{3\pi^2m_a^2f_a^2\lambda^2}
\end{equation}
with $E_a\equiv E_0\cos(m_a t)$ ( $E_0=-k_a\alpha a_0B/f_a\pi=-g_{a\gamma\gamma}a_0B$ ),
where we have used the formula of the London penetration depth $\lambda=\sqrt{m_c/q^2n}$. We note that the dipole current is $I\equiv 2\pi R\lambda J$.
Numerically, we estimate the flux $S$,

\begin{equation}
\label{NS}
S\simeq 4.1\times 10^{-18}\mbox{W}\Big(\frac{5\times10^{-6}\rm cm}{\lambda}\Big)^2\Big(\frac{B}{3\rm T}\Big)^2
\Big(\frac{R}{1\rm cm}\Big)^2\Big(\frac{l}{10\rm cm}\Big)^2\Big(\frac{k_a}{1.0}\Big)^2\Big(\frac{\rho_a}{0.3\rm GeV/\rm cm^3}\Big),
\end{equation}
where there is no dependence on the axion mass.
( As an actual material used in superconducting magnet, the superconductor Nb$_3$Sn shows the penetration depth $\lambda\sim 8\times 10^{-6}$cm.
Then, the radiation flux similar to $S$ in the above can be obtained by using $B\sim 5$T. )

\vspace{0.1cm}
The above formula is derived by assuming the QCD axion, namely, by assuming the equality $m_af_a\simeq 6\times 10^{-6}\rm eV\times 10^{12}$GeV.	
In general axion or axion-like particles, we have

\begin{equation}
\label{GS}
S_g\simeq 2.6\times 10^{-17}\mbox{W}\Big(\frac{g_{15}}{m_6}\Big)^2\Big(\frac{5\times10^{-6}\rm cm}{\lambda}\Big)^2\Big(\frac{B}{3\rm T}\Big)^2
\Big(\frac{R}{1\rm cm}\Big)^2\Big(\frac{l}{10\rm cm}\Big)^2\Big(\frac{\rho_a}{0.3\rm GeV/\rm cm^3}\Big),
\end{equation}
where $g_{15}\equiv g_{a\gamma\gamma}/(10^{-15}\rm GeV^{-1})$ and  $m_6\equiv m_a/(10^{-6}\rm eV)$.
When we use the standard models of the QCD axion, $g_{15}$ is given such that 
$g_{15}\simeq 0.14\,m_6 \,( -0.39\,m_6) $ for DFSZ ( KSVZ ) model.
The general formula $S_g$ can be used for the estimation of the detection sensitivity discussed below.

\vspace{0.1cm}
Radiations from normal conducting wire ( cylindrical metal with small radius ) has been proposed in a previous reference\cite{wire}. 
Oscillating current in the metal with finite size arises from 
oscillating surface charge. The surface charge is the charge induced in upper and down surfaces
of the metal, which screens the electric field $E_a$ inside the metal.
The current does not depend on electron density
and is much smaller than the one in the superconductor.
Thus, the radiation flux is extremely smaller than that in the superconductor.

\vspace{0.1cm}
The spectrum of the radiation shows a sharp peak at the frequency $m_a/2\pi$ with the bandwidth $\Delta \omega\sim 10^{-6}m_a/2\pi$.
We can make the flux $S$ to
increase more by enlarging the radius and length of the cylindrical superconductor as well as the strength of the magnetic field $B$. 
The characteristic contribution in the superconductor comes from the penetration depth $\lambda$. Thus, by choosing sample of superconductor with 
smaller penetration depth than $5\times 10^{-6}$cm, we can obtain much larger radiation flux than $S$ in eq(\ref{NS}).

The flux is obtained by integrating a Poynting vector over the sphere with radius $r\gg 2\pi/m_a$ 
around the superconductor; $S=\int S(\theta, r) r^2 d\Omega=\int S(\theta, r)r^2 \sin\theta d\theta d\phi$,
where 

\begin{equation}
S(\theta, r)=\frac{m_a^2 (2\pi Rl\lambda J)^2(\sin\theta)^2}{8\pi r^2} ,
\end{equation}  
where we have taken the polar coordinate.

The dipole radiation is emitted mainly toward the direction ( $\theta=\pi/2$ ) perpendicular to the electric current $I$ flowing in $z$ direction.
Thus, when we measure the radiation
in the direction, we receive relatively strong flux density.  
For example, we observe the radiation using the radio telescope which is a parabolic dish antenna with
diameter $32$m, e.g. Yamaguchi 32-m radio telescope of National Astrophysical  Observatory of Japan.  
We put the cylindrical 
superconductor $100$m away from the telescope. Then,
the observed flux per frequency $P$ is given by

\begin{equation}
P\equiv \int \frac{S(\theta, r)}{\Delta \omega} r^2 d\Omega=\int^{\pi/2-\delta}_{\pi/2+\delta} \frac{S(\theta, r)}{\Delta \omega} r^2\sin\theta d\theta d\phi\simeq 
\frac{m_a^2 (2\pi Rl\lambda J)^2\delta^2}{8\Delta \omega}=\frac{3S\delta^2}{8\Delta \omega}\simeq 0.4\times 10^{-22}\rm W/Hz,
\end{equation}
with $\delta \simeq 16\rm m/100\rm m=0.16$ and $\Delta\omega=10^3$Hz$\big(m_a/(6\times 10^{-6}\rm eV)\big)$.
Thus, the antenna temperature is approximately $T_a\equiv P\eta\simeq 1.5$K with the unit $k_B=1$,
assuming the antenna efficiency $\eta \simeq 0.6$. Therefore, the radiation can be observed
with the radio telescope with diameter such as $32$m. 

\vspace{0.1cm}
We now estimate the detection sensitivity. When we observe the radiation over time $t$ with bandwidth $\delta \omega$, 
the ratio of signal to noise is given by $S/N=\frac{S'_g}{T_{sys}}\sqrt{t/\delta \omega}$ where $S'_g=3\delta^2S_g/8\sim 10^{-2}S_g$ denotes
the radiation flux received by the telescope $100$m away from the superconductor and
$T_{sys}$ is the system noise temperature. 
For instance, $T_{sys}=40$K for Yamaguchi 32-m radio telescope.
Therefore, we find that

\begin{equation}
\label{sen}
\frac{S}{N}\sim 40\times \sqrt{\frac{1\rm MHz}{\delta \omega}}\sqrt{\frac{t}{1\,\rm s}}
\Big(\frac{g_{15}}{m_6}\Big)^2\Big(\frac{5\times10^{-6}\rm cm}{\lambda}\Big)^2\Big(\frac{B}{7\rm T}\Big)^2
\Big(\frac{R}{2\rm cm}\Big)^2\Big(\frac{l}{20\rm cm}\Big)^2\Big(\frac{\rho_a}{0.3\rm GeV/\rm cm^3}\Big)
\end{equation}
with $T_{sys}=40$K, where we have taken the physical parameters $B=7$T, $R=2$cm and $l=20$cm to have 
better detection sensitivity. Note that $g_{15}\equiv g_{a\gamma\gamma}/(10^{-15}\rm GeV^{-1})$ and  $m_6\equiv m_a/(10^{-6}\rm eV)$.
Thus, even for DFSZ axion ( $(\frac{g_{15}}{m_6})^2\simeq 0.1$ ), we reach the sensitivity $S/N \sim 4$ 
when we observe the radiation over $1$ second 
with the bandwidth $\delta\omega=1$MHz. The detection sensitivity does not depend on the QCD axion mass so that
it can be applied to the search of the QCD axion with any masses, keeping the sensitivity.
It can be also applied to the search of axion-like particles. 
Hence, it turns out that our method for the detection of the dark matter axion is very efficient. 

\vspace{0.1cm}
When we use existing radio telescopes, the frequencies we can observe are limited. Probably, 
they are in the range $0.1$MHz $\sim 1000$GHz. The system noise temperatures $T_{sys}$ of
the telescopes are about typically $100$K. If we intend to detect radiations with frequencies lower or higher than those in the range,
we need to make a sensitive receiver for the radiations with the frequencies.

\vspace{0.1cm}
Although we have discussed the observation using the large radio telescope, the identical flux $S'_g$ ( $P$ ) can be obtained even when
we use a small radio telescope. For instance, we may use a radio telescope 
with small diameter $32$cm, by putting the superconductor $1$m away from the telescope. 
Then, the identical flux can be obtained
as long as the dipole approximation holds;
the distance
$1$m much larger than the wave length of the radiation; $1\mbox{m} \gg 2\pi/m_a$.
Although our discussions depend on the dipole approximation, 
the approximation is not necessarily required. It is adopted for simplicity.

The advantage of our proposal is that we can simultaneously search wide bandwidth of the radio frequency without tuning 
the shape of the superconductor. Furthermore, the high detection sensitivity is obtained 
because the radiation flux is large compared with that in resonant cavity experiments. 
In this way, we can observe the radiation from the dark matter axion with radio telescope.  

\vspace{0.1cm}
Here, 
we explain why we obtain much larger flux in our apparatus than flux obtained in resonant cavity.
For comparison, we write down the flux\cite{sikivie} of TM mode in the resonant cavity, 

\begin{equation}
\label{9}
S_{cavity}=\frac{g_{a\gamma\gamma}^2B^2\rho_a VC}{m_a}\times min(Q_l, Q_a)
=\frac{4(k_a\alpha B)^2 V\rho_a}{\pi^2 m_a^2 f_a^2 m_a R_c^2}\times min(Q_l, Q_a)
\end{equation}
with $C=4/(m_aR_c)^2$, the volume $V=\pi R_c^2 l_c$ of the cavity and $min (Q_l, Q_a)= Q_l ( Q_a )$ for $Q_l<Q_a ( Q_a<Q_l)$,
where $R_c$ ( $l_c$ ) denotes the radius ( length ) of the cavity.  $Q_a\sim 3\times 10^6$ and $Q_l=R_c/\delta_c$ with 
the skin depth $\delta_c$ of the cavity; e.g. $\delta_c\sim 2\times 10^{-4}$cm for the radio frequency $m_a/2\pi=1$GHz in copper. 
We note that the resonance condition of the cavity gives $m_a R_c\sim 2.4$ in eq(\ref{9}), i.e. $C\simeq 0.7$.
Then,
$R_c\simeq 11.5$cm for $m_a/2\pi=1$GHz. 
On the other hand, the flux received by the telescope in the above is given by

\begin{equation}
\label{10}
S_{tel}=\frac{3S\delta^2}{8}=\frac{3(k_a\alpha B)^2(2\pi Rl)^2\rho_a}{24\pi^2m_a^2f_a^2\lambda^2}\times \delta^2
=\frac{(k_a\alpha B)^2V_t \rho_a}{2\pi^2m_a^2f_a^2R}\frac{l R}{\lambda^2}\times \delta^2
\end{equation}
with $\delta\simeq 0.16$ and the volume $V_t\equiv \pi R^2l$ of the cylindrical superconductor.
The small factor $3\delta^2/8$ is the solid angle per $4\pi$ when looking at the telescope from the superconductor.
Thus, the ratio of $S_{tel}$ to $S_{cav}$ is 

\begin{equation}
\frac{S_{tel}}{S_{cav}}\simeq \frac{1.2V_tR_c}{4VR}\times \frac{lR}{\lambda^2}\times \frac{\delta^2}{min(Q_l,Q_a)=Q_l}
=\frac{1.2V_t}{4V}\times \frac{R_c}{R}\times \frac{lR}{\lambda^2}\times \frac{\delta_c}{R_c}\times \delta^2,
\end{equation}
where $min(Q_l,Q_a)=Q_l$ with $R_c=11.5$cm and the spin depth $\delta_c=2\times 10^{-4}$cm of copper 
for the radio frequency $1$GHz. The ratio depends only on the parameters ( length, skin depth, penetration depth, e.t.c. ) of each apparatus.
As expected, it does not depend on the magnetic field $B$ or axion photon coupling $g_{a\gamma\gamma}$.

\vspace{0.1cm}
We suppose the radio frequency $1$GHz from the axion with $m_a\simeq 4\times 10^{-6}$eV. 
Numerically, for example, for $V\simeq 50$L with $R_c=11.5$cm and $l_c=120$cm as the cavity parameters, 
we have 

\begin{equation}
\frac{V_t}{V}\simeq 6\times 10^{-4},\,\, \frac{R_c}{R}=11.5, \,\, 
\frac{lR}{\lambda^2}=4\times 10^{11}, \,\, \frac{\delta_c}{R_c}\simeq1.7\times10^{-5} \quad \mbox{and} \quad \delta^2\simeq 2.6\times 10^{-2}.
\end{equation}

Thus, we find the ratio $S_{tel}/S_{cav}\sim 360$.  The large flux $S_{tel}$ mainly comes from the small penetration depth 
$\lambda=5\times 10^{-6}$cm of the magnetic field,
in other words, the large number density of Cooper pairs. ( The main contribution comes from the factor $lR/\lambda^2\simeq 4\times 10^{11}$. )
That is, the difference in the fluxes arises from
the difference between the penetration depth $\lambda$ and the skin depth $\delta_c$.
Therefore, our apparatus for the detection of the axion works more efficiently than resonant cavities do.

\vspace{0.1cm}
We have used the standard formula for the flux $S$ of the dipole radiation. Here we would like to derive the formula in a microscopic point of view.
Then, we will find the applicability of the formula to what the extent the axion mass can be explored. 
As we have pointed out, a Cooper pair oscillates harmonically under the electric field $E_a$
according to the equation of motion, $m_c\dot{v}=qE_a(t)$. The amplitude $qE_0/(m_cm_a^2)$ ( $E_a= E_0\cos(m_at)$ ) of the oscillation is
much small such as $\sim 10^{-18}\mbox{cm}(B/3T)(m_a/10^{-6}\rm eV)^2$. The Cooper pair emits a dipole radiation with the wave length 
$30\mbox{cm}
(6\times 10^{-6} \mbox{eV}/m_a)$ and flux $\dot{w}=q^4E_0^2/3m_e^2$. When the number density of the Cooper pair is $n$,
the number of the Cooper pairs emitting the radiation is given by $N(l)=2\pi R\lambda l n$.
The point is that their emissions are coherent when the wave length $2\pi/m_a$ is larger than the length $l$ of the cylindrical superconductor. Thus, 
we find that the total flux $\dot{w}N^2$ is  
just equal to $S$ shown in eq(\ref{S}),
\begin{equation}
S=\dot{w}N(l)^2=\frac{q^4E_0^2N(l)^2}{3m_e^2}=\frac{m_a^2 (2\pi Rl\lambda J)^2}{3}
\end{equation}
with $J=\frac{q^2E_0 n}{m_am_c}$.
When the wave length $2\pi/m_a$ is shorter than the length $l$, the formula need to change.
For example,
when we explore the axion mass such as $m_a=10^{-4}$eV, 
the coherent radiations arise from the partial region with length $2\pi/m_a\simeq 1.3$cm in the superconductor. Thus, 
the total flux is given such that 

\begin{equation}
S=\dot{w}N(2\pi/m_a=1.3\mbox{cm})^2\Big(\frac{l=10\mbox{cm}}{2\pi/m_a=1.3\rm cm}\Big)\sim 0.5\times 10^{-18}W
\end{equation}
because there are $(10\rm cm/1.3\rm cm)$ regions in each of which the radiations are coherent, but the radiations from different regions are incoherent.


\vspace{0.1cm}

We make a few comments. First, normal, not superconducting currents also flow in the surface. 
The amount of the normal current is proportional to the number density of normal
electrons. It is strongly suppressed in much lower temperature than critical temperature separating normal and superconducting states.
They would not contribute to the power $P$ detected with the receiver so much.
Furthermore, the cylindrical superconductor should be Type $2$ because the superconductivity must hold
even under the strong magnetic fields $3$T. Then, the magnetic field penetrates inside of the type 2 superconductor and the magnetic vortices are formed.
The electric fields $E_a$ would be also produced around the vortices. They
induce oscillating electric currents in the vortices, but they do not emit radiations outside the superconductor. 
This is because the vortices are surrounded by the superconducting state. 
The radiations emitted from the vortices cannot go out the superconductor.
The electric field induced around the vortices will be discussed in a paper now prepared.

Secondly, the strong magnetic field $B$ parallel to the direction along the length of the cylindrical superconductor is produced by coils surrounding it.
They should have open space for the dipole radiations to escape outside the coils and reach the telescope. In particular, the open space should be present
in the $\theta=\pi/2\pm \delta $ directions ( $\delta\simeq 16\rm cm/100\rm cm=0.16$ )  perpendicular to the cylindrical superconductor ( magnetic field ). 
That is, the coils are composed of two parts; one covers the upper side of the superconductor and the other one covers the lower side.
Then, there is an open space in the coils through which the radiations can escape from the coils.
Furthermore,
the whole of the apparatus must be cooled. We need to use a glass container for liquid helium so as for the radiation to pass the container and
reach radio telescope.

Finally,
the axion mass for our main interest is about $4\times10^{-6}\rm eV \sim 10^{-4}\rm eV$. 
The corresponding frequencies ( wave lengths ) of the radiations are $1$GHz$\sim 24$GHz ( $30$cm $\sim 1.3$cm ).
The interest comes from our previous prediction $\simeq 7\times 10^{-6}$eV\cite{iwazakifrb2} for the axion mass, 
which has been obtained by the analysis of repeating fast radio bursts.
But our proposal for the detection of the dark matter axion is not restricted to the mass range. 
It can be also applied to much larger or smaller axion masses
than those of our main interest.
We should note that the radiation flux in eq(\ref{NS}) or eq(\ref{GS}) does not depend on the mass of the QCD axion. 
Furthermore, the detection sensitivity in eq(\ref{sen}) is independent of the QCD axion mass.
Therefore, our method for the detection of the axion can be used for the axion with any masses,
although the magnitude of the radiation flux estimated above depends on the our dipole approximation.

When we detect the axion with mass larger than $10^{-4} $eV,
it would be necessary to change the setup in the standard resonant cavity experiments. 
In our proposal, all radiations emitted by the superconductor of the cylindrical shape can be searched without any essential modification.
We may use the setup itself mentioned above, or
we may diminish the radius and length of the superconductor for the dipole approximation to be valid. 
On the other hand,
when we detect the axion with mass smaller than $10^{-6} $eV, we can use the identical setup. But, for example,
when we try to detect the axion mass smaller than $10^{-8}$eV,
we need to put the apparatus more than $100$m away from the radio telescope
in order to receive the flux estimated with the dipole approximation.  
This is because the wave length of the radiation is larger than $100$m.
In this way, we can simultaneously search wide bandwidth of the radio frequency without essentially tuning 
the shape of the apparatus. 
But we should note that when we use existing radio telescopes, the frequencies we can observe are limited.
Probably, the frequencies under the consideration could be easily detected with existing radio telescopes.

\vspace{0.1cm}
In summary, we have proposed a new method for the detection of the dark matter axion. It is to use a superconductor of the cylindrical shape
in which the oscillating superconducting current flows under strong magnetic field. We have shown that
the superconductor can emit sufficiently large amount of the dipole radiation so as for existing radio telescopes to easily detect them.
The detection sensitivity expected in our method is remarkably high.

\vspace{0.2cm}
The author
expresses thanks to Alexander John Millar, Tatsumi Nitta, Izumi Tsutsui and Osamu Morimatsu  for useful comments and discussions.
Especially, he expresses great thank to Yasuhiro Kishimoto for useful comments.
This work was supported in part by Grant-in-Aid for Scientific Research ( KAKENHI ), No.19K03832.



\end{document}